\documentclass{aastex631}

\received{December 29, 2021}
\accepted{January 3, 2022}

\submitjournal{RNAAS}

\shorttitle{2004~JW$_{52}$ is an Ordinary Jupiter Trojan}
\shortauthors{Seccull}

\begin{document}

\title{128383 (2004~JW$_{52}$) is an Ordinary Jupiter Trojan Asteroid}

\author[0000-0001-5605-1702]{Tom Seccull}
\affiliation{Gemini Observatory/NSF's NOIRLab, 670 N. A'ohoku Place, Hilo, HI 96720, USA}



\begin{abstract}

The Jupiter Trojan asteroid 128383 (2004 JW$_{52}$) was recently reported to have optical colors that are incongruous with its dynamical class. New and archival observations show that this is not the case. This is a reminder that we must always rule out the possibility that the Point Spread Function (PSF) of a minor planet is blended with that of a background sidereal source in survey images before its colors in the associated survey catalog can be considered reliable.   

\end{abstract}


\section{Introduction}
\begin{figure}
\includegraphics[scale=1.1]{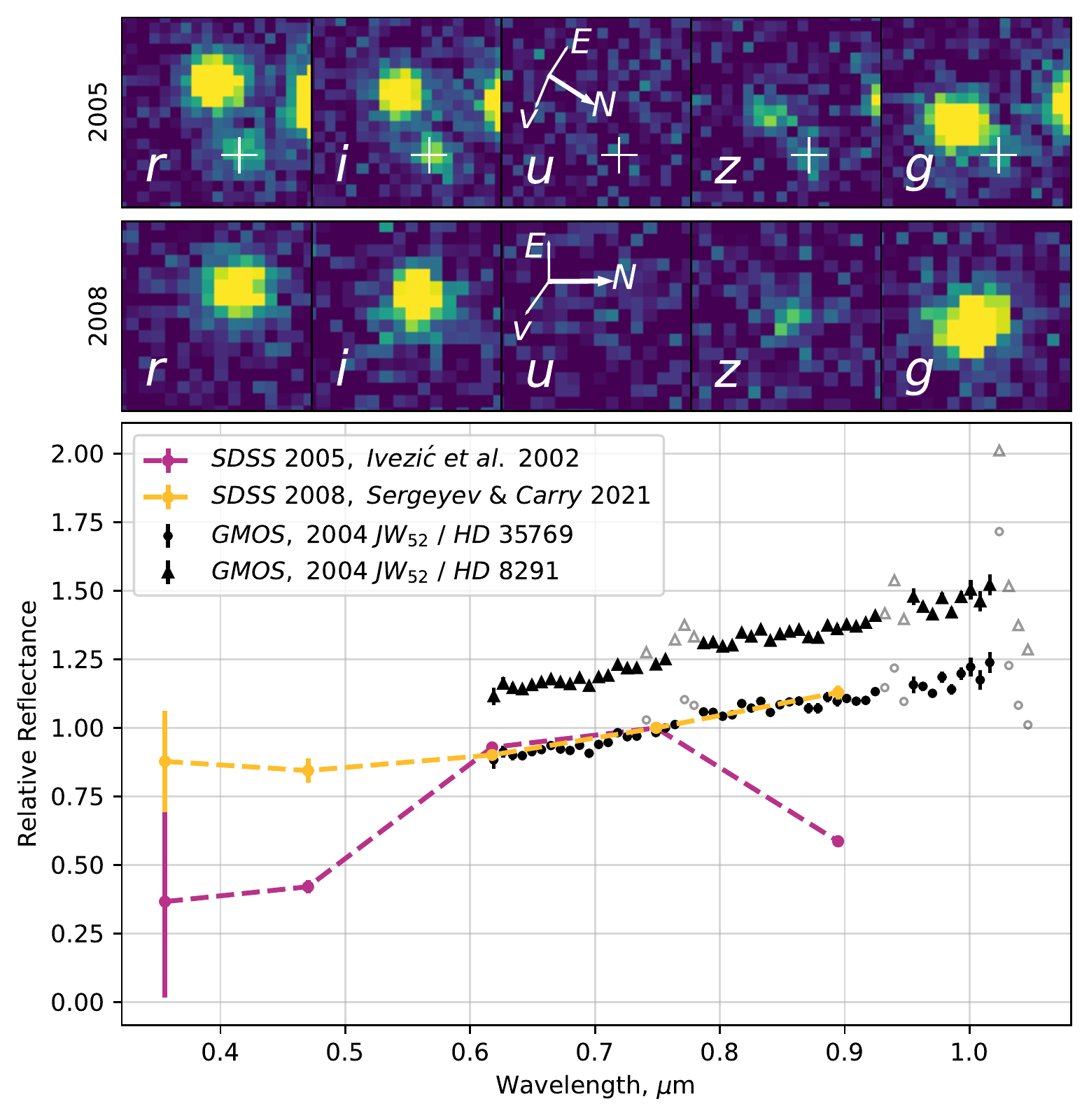}
\caption{Reflectance spectra and SDSS images/colors of 2004~JW$_{52}$. \textbf{Top:} SDSS images of 2004~JW$_{52}$ (the bright central object) observed during SDSS run 5396 in 2005. Time progresses from left to right as the $riuzg$ bands are stepped through in sequence. A white cross marks the background star that 2004~JW$_{52}$ blends with in the $z$ and $g$ images. All sources in these thumbnails are invisible in $u$ band, so in the $u$ panel the average directions of North (N), East (E), and the on-sky motion of 2004~JW$_{52}$ ($v$) during the sequence are plotted. \textbf{Middle:} Same as the top row, but taken during SDSS run 7754 in 2008, when 2004~JW$_{52}$ was clear of background sources. \textbf{Bottom:} The new GMOS-N reflectance spectra of 2004~JW$_{52}$ (each calibrated with a different Solar twin) plotted alongside coarse reflectance spectra derived from the 2005 and 2008 SDSS colors of the same object respectively reported in the catalogs of \citet{2002SPIE.4836...98I} and \citet{2021A&A...652A..59S}. All datasets are scaled to unity at $0.75~\mu$m, with one of the reflectance spectra offset by +0.25 for clarity. Hollow points in the reflectance spectra either fall in the GMOS-N detector chip gaps, or are contaminated with strong sky emission line residuals.
\label{fig}}
\end{figure}

I was intrigued by recent reports that a Jupiter Trojan asteroid, 128383 (2004~JW$_{52}$), has colors in the 4th data release of the Sloan Digital Sky Survey Moving Object Catalog \citep[SDSS MOC;][]{2002SPIE.4836...98I} that are distinct from those of all other Jupiter Trojans recorded there \citep{2021MNRAS.504.1571H}. Conversion of these colors to a coarse reflectance spectrum using the methods of \citet{2002A&A...389..641H}, and SDSS $ugriz$ Solar colors\footnote{\url{https://www.sdss.org/dr12/algorithms/ugrizvegasun/}}, revealed a spectrum more similar to those of S-Complex asteroids, than the featureless red spectra typical of carbonaceous Jupiter Trojans \citep[e.g.][]{2009Icar..202..160D,2011AJ....141...25E}. The strong apparent redness of 2004~JW$_{52}$ at $\lambda<0.6~\mu$m and its strong blueness at $\lambda>0.75~\mu$m suggested that its surface may be silcate-rich \citep[see Fig. \ref{fig}; e.g.][]{2015aste.book...43R}; this would suggest that 2004~JW$_{52}$ originated in the inner Solar System, rather than the primordial Trans-Neptunian disk in which the trojans likely formed \citep[e.g.][]{2020tnss.book...25M}.  Moreover, published dynamical simulations estimate that the resonant capture of 2004~JW$_{52}$ into Jupiter's L4 swarm is stable on gigayear timescales \citep{2020MNRAS.495.4085H}. This stability is difficult to explain for an interloper from the inner Solar System, as Jovian co-orbital asteroids from the main belt are much more likely to be found on unstable horseshoe orbits than they are stable trojan ones \citep{2020AJ....160..144G}. To characterise the surface composition of 2004~JW$_{52}$, constrain its origin, and determine whether it is silicate-rich, I requested Fast Turnaround observations at the 8.1~m Gemini North telescope in 2021 July.

\section{Observations \& Data Reduction}
2004~JW$_{52}$ and two Solar twins \citep[HD~8291, HD~35769;][]{2014A&A...572A..48R,2015A&A...574A.124D} were observed in longslit spectroscopy mode with the Gemini North Multi-Object Spectrograph \citep[GMOS-N;][]{2004PASP..116..425H} on 2021-10-11 UT. A full log of observation geometry and conditions is included with the data behind Figure \ref{fig}. GMOS-N was configured with a $1\arcsec$ wide longslit, the RG610 order blocking filter, and the R400 grating centered at $0.84~\mu$m; this provided a wavelength coverage of $0.61-1.07~\mu$m and a resolving power of $\sim960$ at $0.764~\mu$m. A repeating spatial dither pattern of $0\arcsec, +16\arcsec, +8\arcsec, -8\arcsec$ along the slit was used to enable construction of a fringe frame. During acquisition of each target, the slit was aligned to the average parallactic angle required for the associated spectroscopic observation.

Standard data reduction steps were performed with the Gemini distribution of PyRAF \citep{2012SASS...31..159G}. Cosmic rays were removed with Astro-SCRAPPY \citep{2001PASP..113.1420V}\footnote{\url{https://doi.org/10.5281/zenodo.1482019}}, and I used my own Python3 script to create a fringe frame and subtract fringes from the spatially dithered data. Localisation of the spectrum and sky regions, sky subtaction, spectrum extraction, atmospheric extinction correction, stacking, and calibration for the Solar spectrum was performed as described by \citet{2021PSJ.....2...57S}. The reflectance spectra were binned by a factor of 50 using methods described by \citet{2019AJ....157...88S}.  

\section{Results}
The GMOS-N reflectance spectrum of 2004~JW$_{52}$ (see Fig. \ref{fig}) shows no notable differences when calibrated with either Solar twin, and in both cases is red, featureless, and completely typical of an ordinary Jupiter Trojan asteroid. The gradient of the spectrum, measured with the bootstrapping method of \citet{2019AJ....157...88S}, is found to be $S'=8.0\pm0.5~\%/0.1\mu$m at $0.61-0.92~\mu$m; this places 2004~JW$_{52}$ in the red class of Jupiter Trojans \citep[e.g.][]{2008A&A...483..911R}. In late October I learned of the updated SDSS MOC published by \citet{2021A&A...652A..59S}, and found that colors from another SDSS observation of 2004~JW$_{52}$ from 2008 were completely consistent with the GMOS-N spectrum (see Fig. \ref{fig}). Finally, I retrieved all images of 2004~JW$_{52}$ from the SDSS imaging catalog with the Canadian Astronomy Data Centre's (CADC's) Solar System Object Search tool \citep{2012PASP..124..579G}. Therein I found that the PSF of 2004~JW$_{52}$ had been blended with a background star when it was observed in $g$ and $z$ bands in 2005, and that the photometry in those bands at that epoch were spoiled (see Fig. \ref{fig}). It seems that \citet{2021MNRAS.504.1571H} missed this fact when they remarked on the apparently unusual color properties of 2004~JW$_{52}$; a warning about the possible presence of contaminated photometry in the 4th release of the SDSS MOC is presented by \citet{2002SPIE.4836...98I} in the associated catalog notes\footnote{\url{http://faculty.washington.edu/ivezic/sdssmoc/sdssmoc.html}}. Admittedly, however, I also missed this warning in my eagerness to chase a potentially interesting object with spectroscopy. We should all remember to check the catalog images of an object with unusual colors to make sure it is not blended with background sources before we consider the associated catalog colors to be reliable. This will be of particular importance to those who in future will be looking for unusual minor planets in the data stream of the Rubin Observatory.

\begin{acknowledgments}
 We are grateful for the privilege of observing the Universe from Maunakea, a place that is unique in both its astronomical quality and its cultural significance. Thanks to Ted Rudyk and Sunny Stewart for performing the GMOS-N observations, which were obtained under program GN-2021B-FT-202 at the international Gemini Observatory, a program of NSF’s NOIRLab, which is managed by AURA and the NSF on behalf of the Gemini Observatory partnership. T.S. is supported through a Gemini Science Fellowship. This research made use of SDSS imaging data (\url{https://www.sdss.org/}), NASA's ADS Bibliographic Services, JPL Horizons (\url{https://ssd.jpl.nasa.gov/horizons.cgi}), and the facilities of the CADC. RTFM is a universal and timeless aphorism.
\end{acknowledgments}

%

\vspace{5mm}
\facilities{Gemini(GMOS-N)}


\software{astropy \citep{2013A&A...558A..33A,2018AJ....156..123A}, astroscrappy \citep[\url{https://doi.org/10.5281/zenodo.1482019};][]{2001PASP..113.1420V} Gemini PyRAF \citep{2012SASS...31..159G}, Matplotlib \citep{2007CSE..9..90H}, NumPy \citep{harris2020array}, SciPy \citep{2020SciPy-NMeth}, }


\bibliography{masterbib}{}
\bibliographystyle{aasjournal}



\end{document}